\documentclass[%
 aip,
 amsmath,amssymb,
 reprint,%
]{revtex4-1}

\usepackage{graphicx}
\usepackage{dcolumn}
\usepackage{bm}

\usepackage[utf8]{inputenc}
\usepackage[T1]{fontenc}
\usepackage{mathptmx}
\usepackage{etoolbox}
\usepackage{siunitx}
\usepackage{svg}
\usepackage{amsmath}
\usepackage[usenames,dvipsnames,svgnames]{xcolor} 

\DeclareUnicodeCharacter{2032}{\ensuremath{^{\prime}}}

\makeatletter
\def\@email#1#2{%
 \endgroup
 \patchcmd{\titleblock@produce}
  {\frontmatter@RRAPformat}
  {\frontmatter@RRAPformat{\produce@RRAP{*#1\href{mailto:#2}{#2}}}\frontmatter@RRAPformat}
  {}{}
}%
\makeatother
\begin{document}

\preprint{AIP/123-QED}

\title[1T'-MoTe$_2$ as an integrated SA]{1T'-MoTe$_2$ as an integrated saturable absorber for photonic machine learning}

\author{Maria Carolina Volpato}
\affiliation{Gleb Wataghin Institute of Physics, Universidade Estadual de Campinas (UNICAMP), Campinas, SP, Brazil.}
\author{Henrique G. Rosa}
\affiliation{School of Engineering, Mackenzie Presbyterian University, São Paulo, SP, Brazil.}
\author{Tom Reep}
\affiliation{Photonics Research Group, Department of Information Technology (INTEC), Ghent University–imec, 9052 Ghent, Belgium.}

\author{Pierre-Louis de Assis}%

\author{Newton Cesário Frateschi}
 \email{fratesch@unicamp.br}
 \affiliation{Gleb Wataghin Institute of Physics, Universidade Estadual de Campinas (UNICAMP), Campinas, SP, Brazil.}

\date{\today}

\begin{abstract}
We investigate the saturable absorption behavior of a 1T'-MoTe$_2$ monolayer integrated with a silicon nitride waveguide for applications in photonic neural networks. Using experimental transmission measurements and theoretical modeling, we characterize the nonlinear response of the material. Our model, incorporating quasi-Fermi level separation and carrier dynamics, explains these behaviors and predicts the material's absorption dependence on the carrier density. Furthermore, we demonstrate a coupling efficiency of up to \SI{20}{\percent} between the 1T'-MoTe$_2$ monolayer and the silicon nitride waveguide, with saturation achievable at input powers as low as a few \SI{}{\uW}. These results suggest that 1T'-MoTe$_2$ is a promising candidate for implementing nonlinear functions in integrated photonic neural networks. 
\end{abstract}

\maketitle

Artificial neural networks (ANNs) have revolutionized fields such as image recognition, natural language processing, and decision making by using deep learning techniques to process large amounts of data \cite{lecun2015deep,silver2016mastering,mnih2015human}. However, conventional electronic hardware, rooted in von Neumann architectures, faces significant limitations in running ANNs with the required speed and energy efficiency, fault tolerance, and parallel processing  \cite{indiveri2011neuromorphic,merolla2014million}. Although recent developments in hardware, such as graphics processing units (GPUs), application-specific integrated circuits (ASICs), and hybrid optical-electronic systems, have enhanced performance, these architectures still face scalability issues for the massive parallel computations required by neural networks \cite{benner2005exploitation}. 
With recent advances in silicon photonics \cite{almeida2004all,wang2020integrated}, optical computing has been introduced as an attractive platform to carry out large scale computational schemes. Fully optical neural networks (ONNs) present a promising alternative to their electronic counterparts, offering the potential for unprecedented improvements in computational speed and power efficiency \cite{cardenas2009low,prucnal2017neuromorphic}. Optical systems are highly efficient at executing linear operations, such as matrix multiplications\cite{bogaerts2020programmable}, which forms the backbone of neural network algorithms. However, to implement the essential nonlinear activation functions in these networks, an additional nonlinear component is required \cite{shastri2021photonics}. 

One promising approach involves the use of ring resonators, which can introduce the necessary nonlinearity by leveraging the resonance effects within the optical circuits \cite{basani2024all,garcia2024self}. Alternatively, saturable absorbers have been proposed as another viable nonlinear element to perform activation functions, taking advantage of their intensity-dependent absorption properties has been proposed recently \cite{shen2017deep}. Two-dimensional (2D) materials have emerged as excellent candidates for this function, since they can be saturable absorbers integrated into photonic circuits \cite{bao2009atomic,chen2017transition,chen2024ultra}. In particular, the 1T' phase of molybdenum ditelluride (1T'-MoTe$_2$) has shown great promise, due to its low saturation intensity and high modulation depth \cite{yu2021high}, making it ideal for use in photonic systems that require compact and efficient nonlinear elements. Unlike the semiconducting 2H phase, 1T′-MoTe$_2$ exhibits a gapless band structure, which leads to a high density of available electronic states and fast carrier dynamics \cite{song2018few}.

In this work, we explore the use of 1T'-MoTe$_2$ as a saturable absorber for integrated photonics. To this end, we first characterized the nonlinear response of a monolayer by placing it on the tip of an optical fiber. We propose a model for saturable absorption of this material based on the absorption reduction caused by the quasi-Fermi level separation between electrons and holes as carriers are generated by optical absorption. Using this model and the experimental data, we determine the optical absorption dependence on carrier density which combined with simple carrier dynamics, allowed us to propose and implement experimentally an integrated saturable absorber device using this material on top of a silicon nitride waveguide.  Importantly, the nonlinear characterization of the integrated device was performed using a continuous-wave (CW) laser at telecom wavelengths. This method has been employed for probing nonlinear optical responses in integrated structures such as waveguides and resonators \cite{carmon2007visible,levy2011harmonic,chen2017enhanced,chen2024ultra}. These results revealed the saturable absorption behavior, a key feature for implementing nonlinear functions in optical neural networks.

\begin{figure}[h!]
    \centering
    \includegraphics[width=0.49\textwidth]{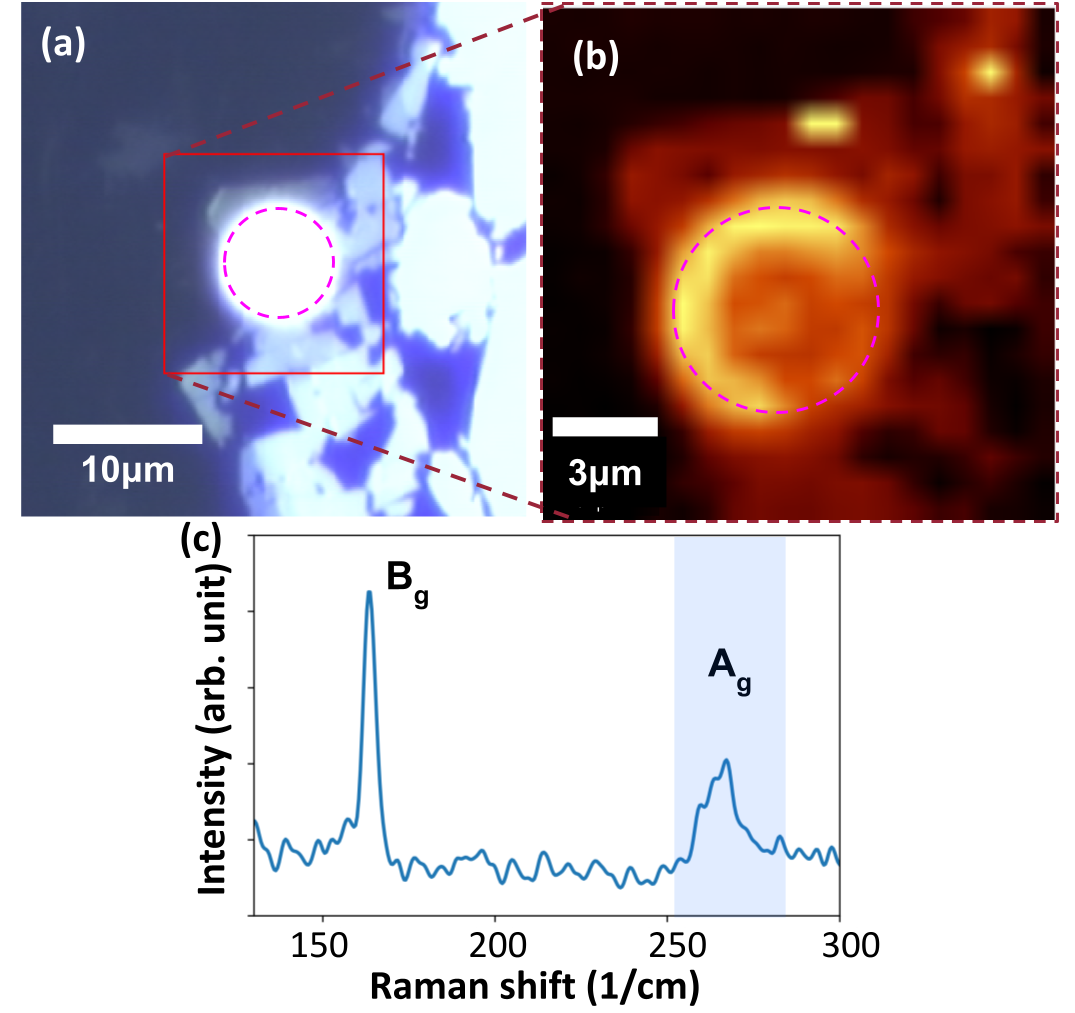}
    \caption{(a) Optical microscopy image of a 1T'-MoTe$_2$ on top of tip of a fiber. (b) Integration of the Raman shift between \SI{247}{cm^{-1}} and \SI{272}{cm^{-1}}, showing that all the fiber core is covered by MoTe$_2$. (c) Raman shift measured at the center of the fiber showing that the flake is a monolayer, based on the characteristic Raman peak position reported in \cite{luo20161t}. The blue region indicates the integration region of (b).}
    \label{fig:raman_fiber}
    \end{figure}

1T’-MoTe$_2$ crystal was obtained by mechanical exfoliation of seed crystal using a low residue tape (Nitto SPV 224) \cite{castellanos2014deterministic,liu2016van,onodera2020assembly}. The procedure is to transfer the flakes onto a PDMS stamp, turn upside down the glass slide with the stamp, align it on top of a target substrate, press the stamp against the substrate, and remove it very slowly. We transferred each selected flake from the PDMS stamp onto the polished face of the optical fiber, as shown in Figure \ref{fig:raman_fiber} (a), via adapted XYZ positioning stages and an optical microscope. This image was obtained coupling a white LED at the fiber, so the bright circle in the center of the image corresponds to the fiber core.

Figure \ref{fig:raman_fiber}(b) shows a Raman map of the fiber core region, obtained by integrating the spectra from \SI{250}{cm^{-1}} to \SI{268}{cm^{-1}}. The spectrum in Figure \ref{fig:raman_fiber}(c) confirms that the transferred 1T'-MoTe$_2$ is a monolayer. These measurements were performed using a confocal Witec Alpha microscope-spectrometer (\SI{532}{nm}, \SI{150}{\micro\watt}), as the $A_g$ peak position correlates with the number of layers \cite{song2018few,VolpatoIPC2023}. Higher counts in the core region arise from silica fluorescence, which adds to the crystal’s Raman signal. The map was also used to estimate the coverage of MoTe$_2$ on the fiber. While a few-layer 1T'-MoTe$_2$ can also function in integrated photonic devices due to its preserved absorption properties, increasing the number of layers leads to higher reflectivity, which reduces the modulation depth \cite{han2021photocarrier,huang2025unveiling}.

\begin{figure}[h!]
    \centering
    \includegraphics[width=0.48\textwidth]{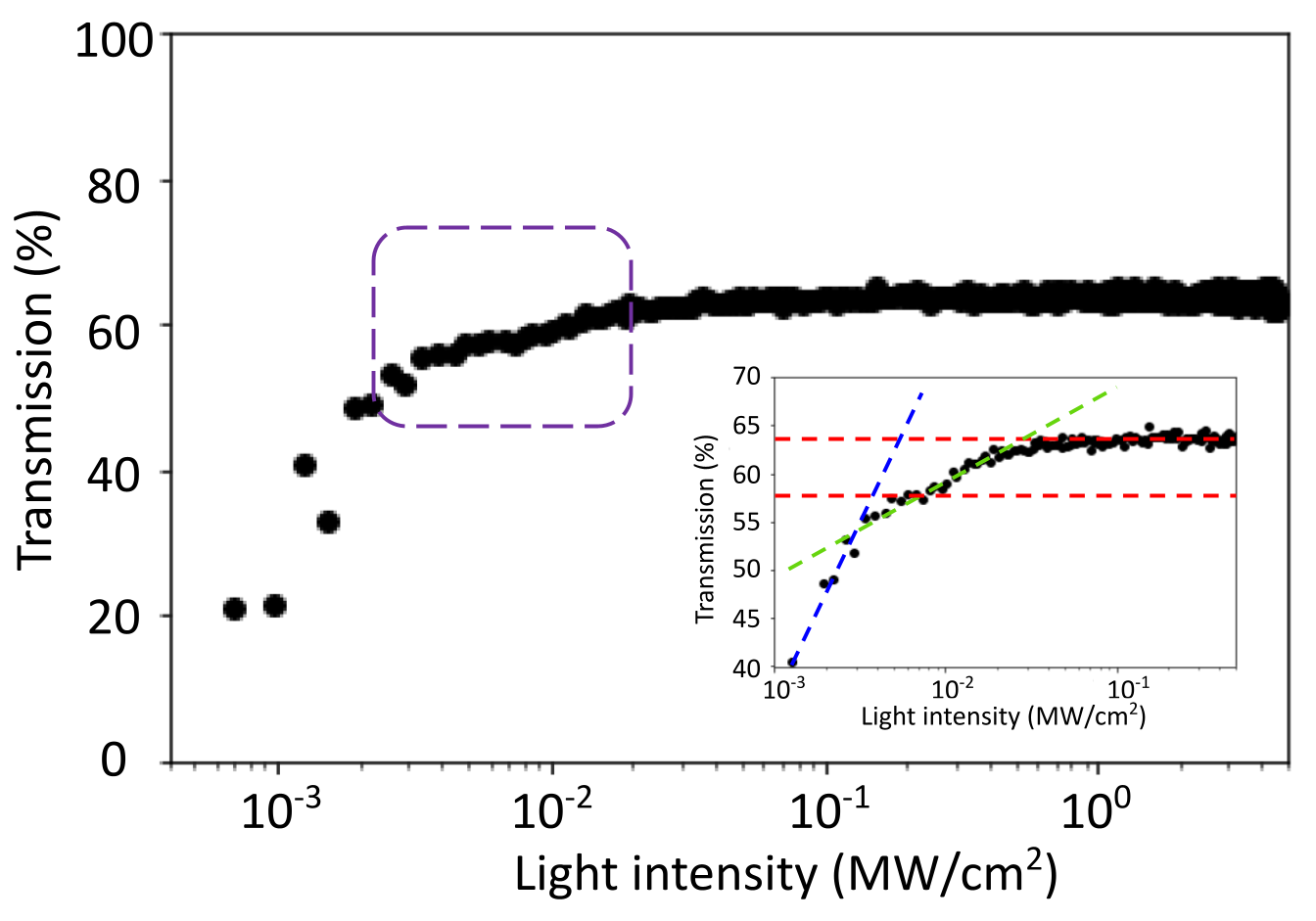}
    \caption{Transmission measurement of a monolayer of MoTe$_2$ on the tip of a fiber using a pulsed laser at \SI{1560}{\nm}, plotted in a semilog scale. Inset: Focus on the region between \SI{0.002}{MW/cm^2} and \SI{0.02}{MW/cm^2} showing a double saturation process.}
    \label{fig:SA}
    \end{figure}

In order to investigate the absorption saturation, we used a pulsed laser source with a center wavelength of $\lambda=$\SI{1560}{nm} and a repetition rate of \SI{89}{MHz}. The pulse duration at the sample was measured using an optical autocorrelator, yielding \SI{2.61}{ps}. This corresponds to a maximum achievable peak intensity of approximately \SI{40}{MW/cm^2}, although in practice the measurements were performed well below this limit, within the few-\SI{}{\MW/cm^2} range where saturation was already observed. The intensity-dependent transmittance was measured using a pair of calibrated fiber-coupled photodetectors, as described in Ref.~\cite{de2022study}. Figure \ref{fig:SA} shows the measured transmission curves where we observe that at low intensities, the transmission increases with increasing intensity until it reaches a plateau.

An important detail shown in the inset of Figure~\ref{fig:SA} is the double-saturation feature in the nonlinear transmission. The first saturation occurs at approximately \SI{2e-3}{MW/cm^2}, while the second appears near \SI{1e-2}{MW/cm^2}. The green dashed line indicates a second saturation not captured by the standard phenomenological model for saturable absorbers, suggesting the presence of two distinct mechanisms. To explain this behavior, we developed a model that accounts for how carrier generation by optical absorption modifies both the absorption coefficient and the reflectivity of the material. First, we calculated the quasi-Fermi levels for electrons and holes under non-equilibrium conditions, which allowed us to determine the absorption coefficient. We also evaluated the change in reflectance due to free-carrier generation upon optical absorption. Our results show that the observed second saturation arises from the interplay between these absorption and reflectance changes.

It is important to note that the saturation intensity obtained for 1T'-MoTe$_2$, approximately \SI{2e-3}{\MW/\cm^2}, is two to four orders of magnitude lower than values typically reported for other 2D materials such as MoS$_2$~\cite{wang2013ultrafast,wang2014broadband}, WSe$_2$~\cite{cao20192}, and black phosphorus~\cite{hisyam2016generation,zhang20192d}. This result highlights the strong light–matter interaction in the semi-metallic 1T' phase of MoTe$_2$, which enables efficient absorption saturation at ultra low optical intensities.

The quasi-Fermi levels are determined from the density of states (DOS) and the Fermi occupation function. Assuming parabolic bands, the DOS in a 2D material is independent of energy and given by $\rho(E) = m / (\pi \hbar^2 \epsilon)$, where $m$ is the effective mass of electrons in the conduction band or holes in the valence band, $\epsilon$ is the material thickness, and $\hbar$ is the reduced Planck constant. Additionally, we assumed that the absorption process does not require a phonon. Unlike the emission process, which requires a lattice phonon, after vertical, no-phonon absorption, the excited electrons relax via intraband-scattering to the bottom of the conduction band. The electron ($n$) and hole ($p$) densities at a given temperature are then obtained from the integrals

\begin{equation}
    n=\int ^\infty _{E_c} \rho(E) f(E) dE=\frac{m}{d\pi \hbar^2}\int ^\infty _{E_c}\frac{1}{e^{(E-E_{fc})/k_B T}+1}dE,
    \label{n(E)}
\end{equation}

\begin{equation}
\begin{split}
    p=\int _{-\infty} ^{E_v} \rho(E)(1- f(E)) dE \\
    =\frac{m}{d \pi \hbar^2}\int _{-\infty} ^{E_v}\frac{1}{e^{(E_{fv}-E)/k_B T}+1}dE, 
    \label{p(E)}
\end{split}
\end{equation}
where $E_v$ is the maximum energy at the valence band, $E_c$ is the minimum energy at conduction band, $E_{fc}$ and $E_{fv}$ are the quasi-Fermi levels for conduction band and valence band respectively. 

\begin{figure}[h!]
    \centering
    \includegraphics[width=0.48\textwidth]{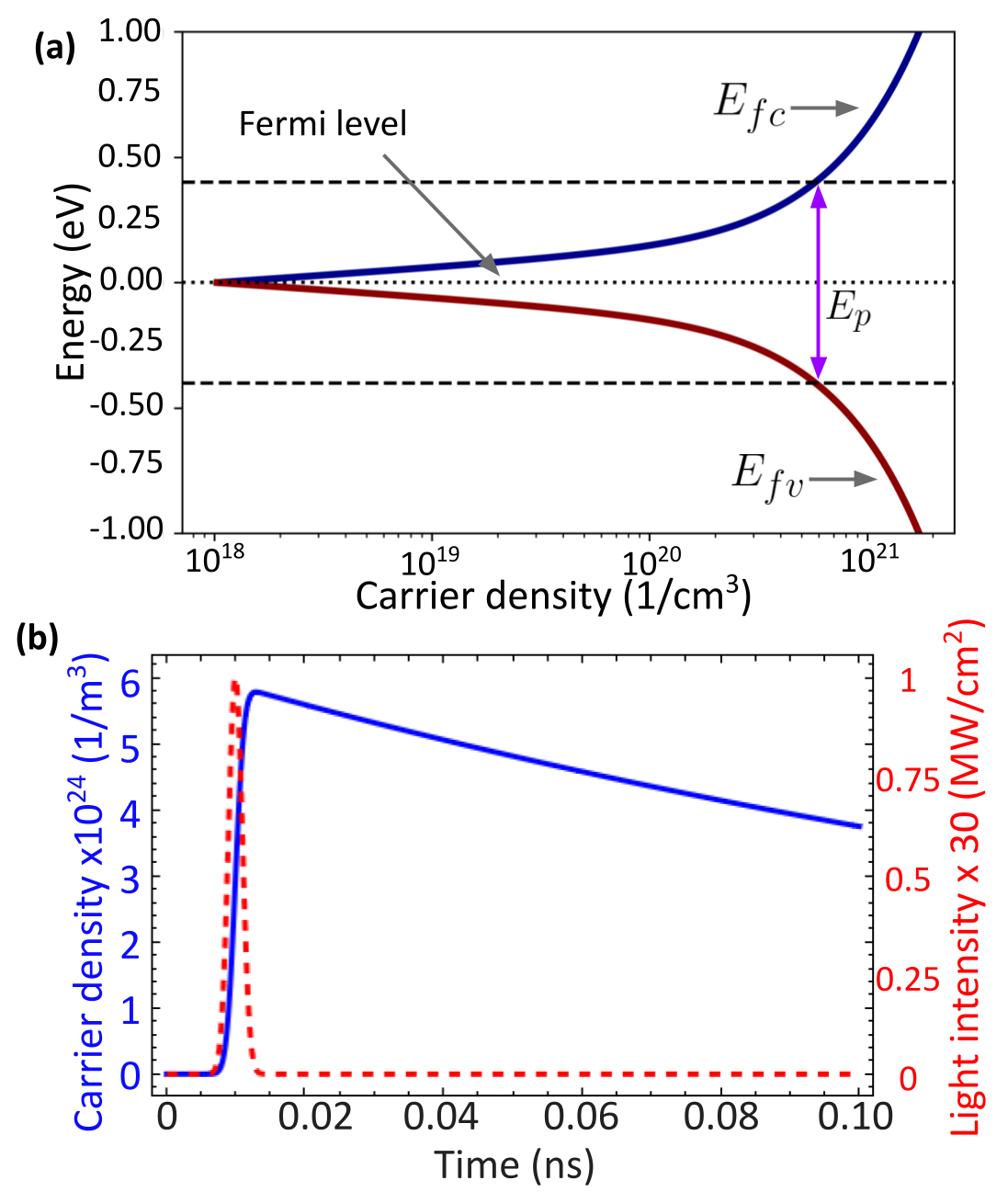}
    \caption{(a) Quasi-Fermi levels calculation using the gap energy ($E_g=E_c-E_v$) $E_g=$\SI{-0.2}{eV} \cite{huang2016controlling}, $m=0.462m_0$ \cite{keum2015bandgap}, the thickness of a monolayer of 1T'-MoTe$_2$ of \SI{1}{nm}. The dashed line represents when the absorption is saturated because the separation of the quasi-Fermi levels is higher than the photon energy $E_p=$ \SI{0.8}{eV} ($\approx$ \SI{1550}{nm}). (b) Carrier density (blue) and laser pulse (red dashed line). For this calculations we used $\tau=$\SI{200}{ps}, $\sigma=$ \SI{2}{ps}, $\alpha=$ \SI{1e6}{1/m}, $I_0=$\SI{30}{MW/cm^2}, $E_p=$ \SI{0.8}{eV} and $\eta=1$, these values are based on the experiment.}
    \label{fig:fig3}
    \end{figure}

In the absorption process of a photon an electron-hole pair is created. Therefore, as we increase the number of photons, carrier generation occurs and the quasi-Fermi level of electrons and holes change. When the separation of the conduction band quasi-Fermi level $E_{fc}$ and the valence band quasi-Fermi level $E_{fv}$ is greater than the energy of the input photon, the material becomes transparent. To understand this process in the MoTe$_2$ we determine the quasi-Fermi level for the valence and the conduction band and obtain the curves solving Eq. \ref{n(E)} and \ref{p(E)} numerically, as shown in Figure \ref{fig:fig3}(a).

The absorption coefficient for a given photon energy $E$ as a function of the quasi-Fermi levels was calculated considering transitions between the conduction band ($E_c$ and $\vec{k_c}$) and the valence band ($E_v$ and $\vec{k_v}$), where electrons (holes) are in quasi-equilibrium with the quasi-Fermi levels $E_{fc}$ ($E_{fv}$). We also consider that there are $N(E)$ resonant states with photon energy between $E$ and $E + dE$ in a volume $V$.

The probability of an electron occupying the $E_c$ level is given by $f_2=1/{e^{(E_c-E_{fc})/k_BT}+1}$ and the probability of an hole occupying the $E_v$ level is given by $f_1=1/{e^{(E_v-E_{fv})/k_BT}+1}$. Using Fermi's golden rule and the relation between absorption and the quasi-Fermi levels \cite{agrawal2000nonlinear}, we can obtain the final expression for the absorption coefficient:
    \begin{equation}
        \alpha=\xi f_2(k_0) \left[1-f_1(k_0) \right]\left(1- e^{(E-(E_{fc}-E_{fv})/k_B T} \right), 
    \end{equation}
    where
    \begin{equation}
    \xi=\frac{q^2 h}{\pi \hbar^2 d v_g m} \left( \frac{1}{\varepsilon E_p}\right)M,
    \end{equation} 
and $k(0)=k_0=\sqrt{m(E-(E_{fc}-E_{fv}))/\hbar^2}$, $q$ is the electron charge, $d$ is the material thickness, $v_g$ is the group velocity, $m$ is the effective mass of electrons and holes, $\varepsilon$ is the permittivity of the material, $E_p$ is the photon energy, and $M$ is a matrix element of the momentum operator between initial and final states.

The carrier dynamics can be written as
\begin{equation}
    \frac{\partial \delta n(x,t)}{\partial t}= G(n(x,t))-\frac{\delta n(x,t)}{\tau}+D\nabla^2 \delta n(x,t),
    \label{eq:dndt}
\end{equation}
where $G(n(x,t))$ is the carrier generation rate, $\tau$ is the carrier lifetime and $D$ is the diffusion coefficient. Although sub-picosecond relaxation components have been reported for 1T'-MoTe$_2$, the effective carrier lifetime relevant to steady-state saturation can be considerably longer, around tens to hundreds of picoseconds \cite{han2021photocarrier,huang2025unveiling}. To a first approximation, for the material at the fiber tip, the carrier density does not depend on the position within the material because the pulse passes through the monolayer in approximately \SI{3e-18}{s}, so we neglect the diffusion term.

The time dependent solution for a Gaussian laser pulse is given by
\begin{equation}
    \frac{d\delta n(t)}{dt}= \frac{\eta \alpha I(t)}{E_p}-\frac{\delta n(t)}{\tau},
\end{equation}
where $ I(t)=I_0 e^{-\frac{1}{2}\left( \frac{t-t_0}{\sigma/2}\right)^2}$, $I_0$ is the maximum intensity, $t_0$ is the time when the center of the pulse coincides with the center of the monolayer, $E_p$ is the incident photon energy, and $\sigma$ is the pulse width. Figure \ref{fig:fig3}(b) shows the laser pulse and the carrier density changing at the center of the monolayer as a function of time, where we can see that the carrier density decays slowly and has a peak dislocated of the pulse peak.

Integrating the carrier density as the pulse passes through the material we can obtain the mean carrier density. And with this value we can use the Drude-Lorentz model \cite{fox2010optical} to determine the refractive index of the monolayer for different pump intensities. For modeling the experiment we have to consider that there is a reflection in the MoTe$_2$/tip of the fiber---which is made of silica---interface. So, the reflectance $R(I)$ of this interface is given by
\begin{equation}
    R(I)= \left(\frac{\bar{n}(I)-\bar{n_1}}{\bar{n}(I)+\bar{n_1}} \right)^2,
    \label{eq:R}
\end{equation}
where $\bar{n}(I)$ is the refractive index of the monolayer and $\bar{n_1}$ is the refractive index of the silica. 

The complete model can be obtained by adding both effects in the transmission calculations, as
\begin{equation}
    T(I)= 1-T_0-e^{-\alpha(I)\,\varepsilon}-R(I),
\end{equation}
where $T_0$ is the transmission at low light intensity and $\alpha(I)$ is the absorption coefficient as a function of the peak intensity of the pulse.

\begin{figure}[h!]
        \centering
        \includegraphics[trim={0 0.1cm 0 0},clip,width=0.45\textwidth]{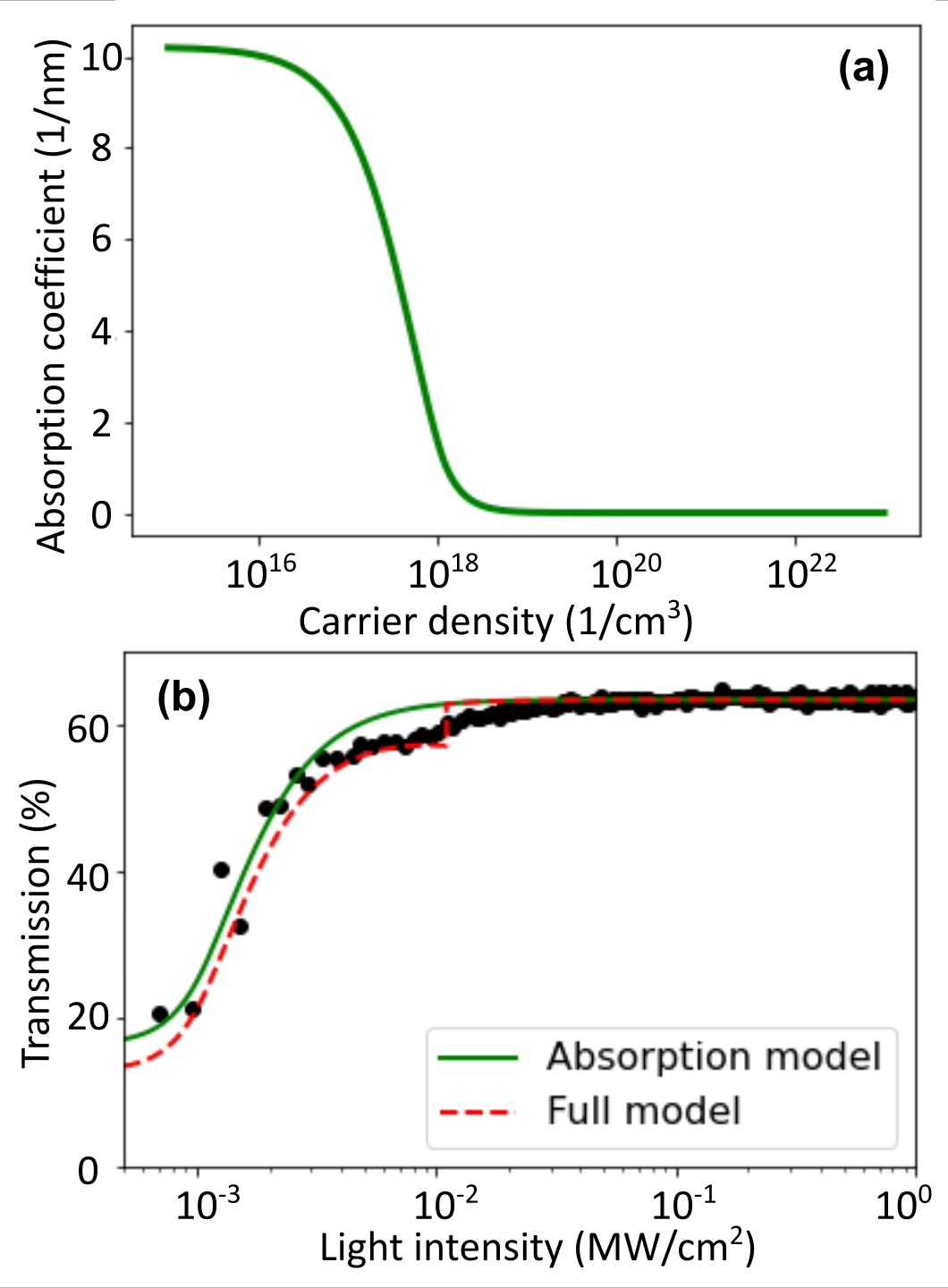}
        \caption{(a) Calculated absorption coefficient as a function of the carrier density. (b) Experimental results in black, calculated transmission considering only the variation of the absorption coefficient in green and calculated transmission considering the variation of the absorption coefficient and the reflectance for different carrier densities. The fitted parameters for the absorption model were $m_{eff}=0.005$ and $\alpha=$\SI{1e10}{1/m}. Reprinted with permission from \cite{volpato2024analysis}.}
        \label{fig:fig4}
    \end{figure}

At low intensities the quasi-Fermi level separation and the effect of carrier dynamics at the reflectance are negligible. So we can obtain the absorption coefficient from the fiber tip measurements at low intensities. Using quasi-Fermi level separation, as obtained from carrier dynamics and Fermi statistics, we then calculated the absorption for different light intensities, as shown in Figure \ref{fig:fig4}(a). Absorption saturation is clearly predicted for higher light intensity.

Figure \ref{fig:fig4} (b) shows the experimental results and the curves fits considering only the variation of the absorption coefficient and a full model considering the absorption coefficient and reflectance for different carrier densities. For this plot, we manually fitted the theoretical curves to qualitatively match the experimental data, indicating that the full model can accommodate both sharp and smooth transitions depending on the chosen parameter set. The effective mass for transport calculations---used in the reflectance calculation---and for the density of states calculation---used for the quasi-Fermi level calculations---were used as fitting parameters. The effective mass for the absorption and reflectance are equal to $m=0.005m_0$ and for $m=m_0$, respectively. The sharp transition in the transmission curve observed in the full model (red dashed line) reflects an abrupt change in absorption with increasing light intensity. In particular, parameters such as the saturation intensity, carrier lifetime, and broadening mechanisms significantly influence the slope of the transition. 

Given the complex structure of the valence and conduction bands, also, considering the 2D nature of the material, the effective masses values related to transport properties are certainly different from those to be employed in a simplified single parabolic band density of states calculation. Also, the effective mass given by electrical measurements is different from the effective mass measured by optical methods \cite{ashcroft1976solid}. For electron transport, the effective mass, derived from the band structure curvature near the conduction band minimum, determines the electron's response to electric fields, affecting mobility and conductivity. Conversely, the optical effective mass involves the combined mass of excitons relevant to optical transitions and is crucial for understanding absorption, emission spectra, and exciton binding energy. Moreover, the density of states for a 2D material requires the determination of the thickness of the layer, which, for a single atomic layer, would require a refined model for the mesoscopic structure. In order to consider all these effects, we have assumed that two sets of value for the effective masses need to be considered, one for transport used for the Drude-Lorentz model, an one for the quasi-Fermi level calculation. Both values were used as fitting parameter.

\begin{figure*}[t!]
        \centering
        \includegraphics[width=0.98\textwidth]{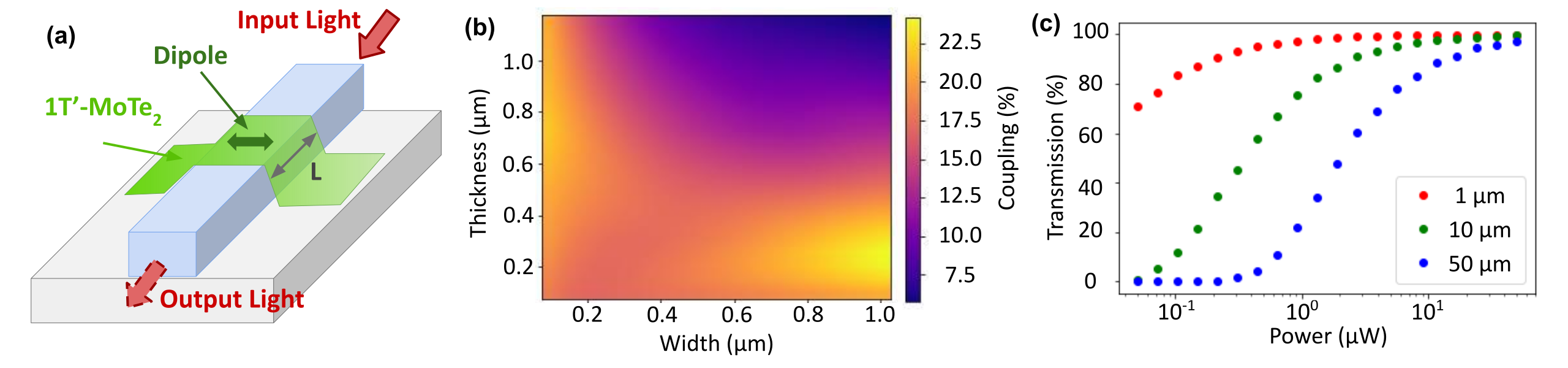}
        \caption{(a) Cross section of the simulated structure with indications: the silicon nitride waveguide in purple, the SiO$_2$ substrate in gray the dipole on top of the waveguide in green. (b) Simulation of coupling the dipole radiation with the SiNx waveguide for different thicknesses. (c) Calculated transmission as a function of pump power for different device lengths. (a) and (c) reprinted with permission from \cite{volpato2024analysis}.}
        \label{fig:fig5}
    \end{figure*}

After studying the behaviour of the 1T'-MoTe$_2$ as a saturable absorber, we proceeded to use the obtained parameters and simulation results to design and then fabricate an integrated device for testing. The proposed device, shown in Figure \ref{fig:fig5} (a), is obtained by placing a monolayer of MoTe$_2$ on top of a waveguide. This structure offers the opportunity to make integrated saturable absorbers compatible with foundry Multi-Project Wafer (MPW)\footnote{At least ANT, Imec and Ligentec offer the process of opening cladding windows to waveguides in their MPW runs}, making it a significant alternative for on-chip saturable absorbers.

For an integrated saturable absorber the wave vector of the incident light is perpendicular to the normal vector of the 2D material, so we will neglect the reflectance of the material. For the steady state, i.e. using a CW laser as input power, the carrier density can be calculated as a function of the intensity at the origin -- which corresponds to one edge of the material-- minus the intensity at position x, that is
\begin{equation}
    \delta n(x)= \frac{\tau \eta }{E_p}  \left(\frac{I(0)-I(x)}{x} \right).
\end{equation}
Assuming $I(x)= I_0 e^{-\alpha(n(x)) x}$, where $n(x)=n_0+\delta n(x)$, we obtain the transmission for each position as
\begin{equation}
    T=\frac{I_{out}}{I_{in}}=\frac{I(L)}{I_0}=  e^{-\alpha(n(L)) L},
    \label{eq:transm}
\end{equation}
where $I_{out}$ is the output light intensity, $I_{in}$ the input light intensity and $L$ is the length of the monolayer.

We calculate the coupling between the 2D material and the optical mode using Finite Difference Time Domain (FDTD) simulations. A transmission monitor was placed \SI{10}{\um} from a dipole positioned at the center of the waveguide, representing the monolayer exciton \cite{liebermeister2015photonic}, to quantify the fraction of emitted light guided by the waveguide. It is important to note that we are examining intralayer carriers, oriented within the plane of the waveguide surface, as depicted in Figure \ref{fig:fig5}(a). Figure \ref{fig:fig5}(b) shows that the maximum coupling for the saturable absorber reached approximately \SI{20}{\percent} for a waveguide \SI{300}{nm} thick and \SI{900}{nm} wide.

Using this \SI{20}{\percent} coupling in Eq.\ref{eq:transm}, we determine the transmission as a function of input intensity for different monolayer lengths. Figure \ref{fig:fig5}(c) illustrates the expected saturation behavior at various pump powers, showing that longer devices require higher powers to reach saturation. For a typical 1T'-MoTe$_2$ membrane of \SI{10}{\um}, saturation can be achieved with only a few \SI{}{\uW} of power.

\begin{figure}[h!]
        \centering
        \includegraphics[width=0.48\textwidth]{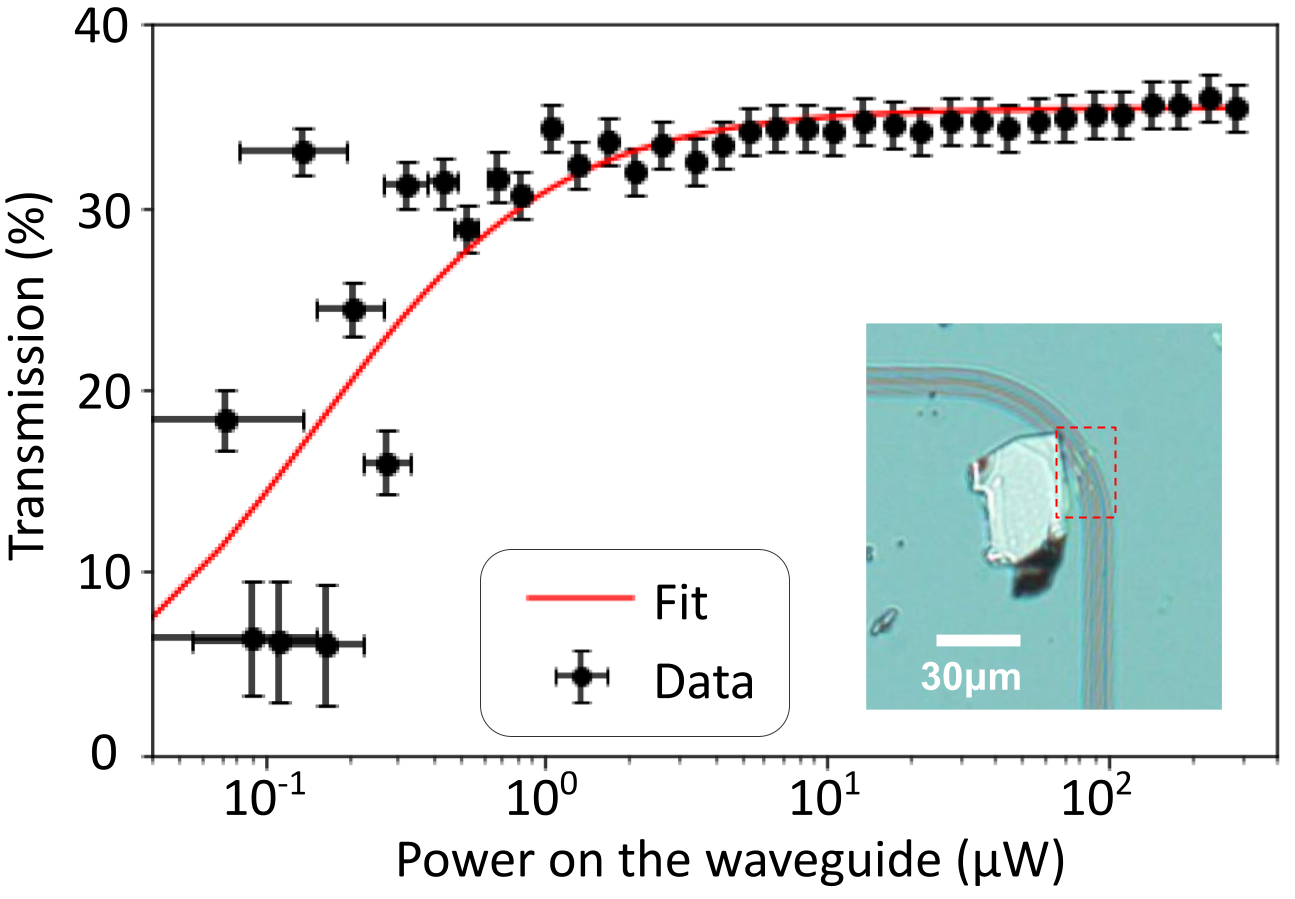}
        \caption{Transmission measurement of a monolayer of 1T'-MoTe$_2$ on top of a silicon nitride waveguide, plotted on a semi-logarithmic scale. The fitted saturation power is \SI{0.15(9)}{\uW}. Transmission values are calculated as the ratio between the measured power after and before the material transfer, representing a relative transmission. The power in the waveguide was estimated based on the measured insertion loss, assuming symmetric coupling losses at the input and output grating couplers. Inset: Optical microscopy image of the 1T'-MoTe$_2$ transferred onto the waveguide.}
        \label{fig:fig6}
    \end{figure}

To validate these predictions experimentally, we transferred a 1T'-MoTe$_2$ monolayer near a bulk flake onto a silicon nitride waveguide patterned by e-beam lithography and RIE etching, as shown in the inset of Figure \ref{fig:fig6}. Figure \ref{fig:fig6} presents the measured saturation response using a CW \SI{1550}{\nm} laser, with incident power varied by a digital variable attenuator. For calibration, the same device was measured before TMD transfer, so the reported transmission values correspond to the transmitted power with the 2D material divided by the transmission in the calibration measurement. This procedure assumes that insertion losses, primarily due to coupling between the fiber and the waveguide via grating couplers, remain approximately constant between measurements.

Fitting the experimental results on Fig. \ref{fig:fig6}, we obtained a saturation power of \SI{0.15(9)}{\uW}, which is in reasonable agreement with a \SI{9}{\um} length 1T'-MoTe$_2$. Additionally, the measured transmission varies from approximately \SI{8}{\percent} to \SI{36}{\percent}, rather than reaching \SI{100}{\percent}. This partial modulation may be caused from different fiber-to-grating alignment before and after the transfer, which can affect the insertion loss estimation. Another possible contributing factor is scattering caused by the bulk portion of the MoTe$_2$ flake located near the waveguide core. Considering the submicron mode area of the silicon nitride waveguide, the corresponding optical intensity is comparable to that observed in the fiber-tip configuration, confirming that true saturable absorption occurs in the integrated platform. The low saturation threshold demonstrates that modest optical powers are sufficient to induce nonlinear absorption, reinforcing the potential of 1T'-MoTe$_2$ as an on-chip nonlinear element—a key requirement for scalable photonic machine learning architectures\cite{shen2017deep}.

We have demonstrated the potential of 1T'-MoTe$_2$ as a saturable absorber integrated with silicon nitride waveguides, specifically targeting applications in photonic neural networks. Through a combination of experimental transmission measurements and theoretical modeling, we investigated the absorption characteristics of 1T'-MoTe$_2$ under varying light intensities. The development of a model incorporating quasi-Fermi level separation and carrier dynamics allowed us to explain the "second" saturation observed experimentally. This model was then extended to predict the performance of an integrated saturable absorber device using 1T'-MoTe$_2$. We showed that the coupling efficiency between the monolayer and the silicon nitride waveguide could reach up to \SI{20}{\percent}, and that saturation could be achieved with only a few \SI{}{\uW} of input power, making it a compact and efficient nonlinear component. 

\begin{acknowledgments}
The authors would like to thank Dr. Ingrid D. Barcelos, Dr. Alisson R. Cadore,  Prof. Paulo Eduardo de Faria Jr. for the fruitful discussions during the preparation of this manuscript and MackGraphe (Center for Research in Graphene and Nanotechnology) at Mackenzie Presbyterian University for providing the facilities and technical support necessary to perform the optical saturation experiments. Conselho Nacional de Desenvolvimento Científico e Tecnológico (CNPQ) through grants No. 440231/2021-3 and No. 305282/2022-0,; Coordenação de Aperfeiçoamento de Pessoal de Nível Superior (Financial Code 001); Fundação de Amparo à Pesquisa do Estado de São Paulo (FAPESP) through grant No. 2018/25339-4, and Air Force Office of Scientific Research (AFOSR) through Grant No. FA9550-20-1-0002.
\end{acknowledgments}

\section*{Disclosures} The authors declare no conflicts of interest.

\section*{Data Availability Statement}
The data that support the findings of this study are available from the corresponding author upon reasonable request.

\bibliography{main}

\end{document}